Looking at Software Sustainability and Productivity Challenges from NSF[1]

Daniel S. Katz and Rajiv Ramnath, National Science Foundation
{dkatz, rramnath}@nsf.gov

The National Science Foundation (NSF) Division of Advanced Cyberinfrastructure (ACI) (within the Directorate for Computer & Information Science, CISE) coordinates and supports the acquisition, development and provision of state-of-the-art cyberinfrastructure (CI) resources, tools and services essential to the conduct of 21st century science and engineering research and education. ACI supports CI resources, tools and related services such as supercomputers, high-capacity mass-storage systems, **system software suites and programming environments, scalable interactive visualization tools, tools that enable scientific collaboration, productivity software libraries and tools**, large-scale data repositories and digitized scientific data management systems, networks of various reach and granularity and **an array of software tools and services** that hide the complexities and heterogeneity of contemporary CI while seeking to provide ubiquitous access and enhanced usability.

Software is an integral enabler of computation, experiment and theory. Scientific discovery and innovation are advancing along fundamentally new pathways opened by development of increasingly sophisticated software. Software is also directly responsible for increased scientific productivity and significant enhancement of researchers' capabilities. In order to nurture, accelerate and sustain this critical mode of scientific progress, NSF has established the Software Infrastructure for Sustained Innovation ($SI^2$) program, with the overarching goal of transforming innovations in research and education into sustained software resources that are an integral part of CI.

$SI^2$ is a long-term investment focused on catalyzing new thinking, paradigms, and practices in developing and using software to understand natural, human, and engineered systems. The intent of $SI^2$ is to foster a pervasive CI to help researchers address problems of unprecedented scale, complexity, resolution, and accuracy by integrating computation, data, networking, observations and experiments in novel ways. It is the expectation of NSF that $SI^2$ investment will result in robust, reliable, usable and sustainable software infrastructure and will transform science and engineering while contributing to the education of next generation researchers and creators of future CI. Education at all levels will play an important role in integrating such a dynamic CI into the fabric of how science and engineering is performed.

Since 2010, $SI^2$ has funded the development and maintenance of about 120 projects, a combination of small projects that support self-contained software elements (called SSEs) and larger framework projects that integrate multiple SSE-level software packages (called SSIs). 4 of these were collaborative US-UK SSI projects in computational chemistry. These projects are of particular interest due to the complexities of international funding that were overcome. In general, most funding agencies are not willing or able to fund work performed outside their borders. This means that international collaborations need to be funded by multiple agencies. However, each agency normally has its own processes and own cycles, which rarely line up with those of other agencies. In this case, the two agencies involved (NSF in the US and EPSRC in the UK) agreed on a common process to fund these projects, which was then successfully carried out. *The lesson from this experience regarding international collaboration is that it is best that the funding agencies agree in advance on their goals and objectives, and that communities of software developers and users may need to work together to convince the agencies to do this.*

SI2 is now beginning the process of institutes, which are intended to address larger community needs rather than simply funding individual software projects, though there is some overlap – software projects often impact their communities beyond just the use of the software, and software institutes will likely also contribute to some software directly. *We believe the process of institution conceptualization awards, which*

---

[1] Please cite as: Katz, Daniel S. and Ramnath, Rajiv. Looking at Software Sustainability and Productivity Challenges from NSF. arXiv.org, 17 August 2015, http://arxiv.org/abs/1508.03348

*bring together communities in planning potential institutes, leading into institution implementation awards will strongly contribute to community-wide awareness of software issues, software development, and software integration, leading to novel advances in science and engineering.*

As the SI2 program has been carried out, we have identified a number of sustainability and productivity challenges. One is **funding models**. NSF generally supports projects for up to five years, but the lifetime of software projects can span twenty or more years. We don't want valuable software to disappear simply because there are no funding sources to support ongoing maintenance and support. Is the answer open source, community-supported software? SI2 has generally been successful in creating a funding model that supports integration, porting, and usability improvement, but continued effort is needed to ensure that the NSF processes and the project reviewers do not slide back to just reviewing project based on their immediate innovativeness. Another challenge is related to **career paths for software-focused researchers**. Since the university structure and academic culture rewards publications, researchers whose main products are software are at a disadvantage. They may feel forced to leave academia for industry to go where their skills are recognized, even if the impact of their work is not in the science area to which they wanted to contribute. Related to this is the challenge of **incentives, including credit**, how the contributors to software are recognized for their contributions, whether in software design, testing, patching, documentation, outreach, etc. There is currently no standard method for publishing or citing software (though a number of projects are working on this), and we already know that the methods we have for publishing papers are beginning to be recognized as failing to identify all the contributions to the papers. Another challenge is **training**. There is a range of types of software developers, from professional software engineers to those end users who also contribute to software. With the possible exception of full-time scientific software developers, there is often a lack of awareness and knowledge on applying software engineering practices, such as in right-sizing software engineering practices to the "micro" teams that make up most of the software developers in science. Another challenge is in **interdisciplinary work**. Many scientists who contribute to software have the ability to do so in multiple scientific and engineering domains, either combining disciplines or working in both computer science and another science or engineering area, but doing so doesn't fit our siloed system and culture and is often discouraged. A sixth challenge is **portability**, or dealing with platform change. This is particularly important today, as we are in a period of flux of hardware and middleware. Developers need to plan for their software to work on different types of hardware, including accelerators. And software written in different languages (e.g., Fortran, Java, Python, Perl, C/C++, Matlab, R) and on different platforms (e.g., cluster, HPC, cloud, grid) frequently must be combined to accomplish important science and engineering objectives. A final challenge is **dissemination**, how scientists know about available software and examples of how it has been used, strengths, weaknesses, and the experience of other users. *The lessons from these experiences are that both funding agency and academic culture do not change as quickly as scientific practices, and that continued effort is needed to raise these issues. We have dealt with some of them internally at NSF, through SI2 as a whole, and specifically, through Institutes, EAGER awards, workshops, including hackathons, and are working towards additional community workshops and national activities intended to create awareness of the problems and of successes and failures in attempting to solve them.*

Overall, we believe that the SI2 program has been very successful to date, but that there is a lot of work remaining, including carrying out the institute process, and both leading and pushing academic community to develop solutions to the existing sustainability and productivity challenges.